\documentclass[twocolumn,showpacs,secnumarabic,amssymb,nobibnotes,aps,prl]{revtex4-1}

\usepackage{amssymb}
\usepackage{multirow}
\usepackage{graphicx}
\begin{document}
\title{Hydrodynamic Effects in Kinetics of Phase Separation in Binary Fluids: Critical versus off-critical compositions}
\author{Koyel Das and Subir K. Das}
\email{das@jncasr.ac.in}
\affiliation{Theoretical Sciences Unit and School of Advanced Materials, Jawaharlal 
Nehru Centre for Advanced Scientific Research,
Jakkur P.O., Bangalore 560064, India.}
\begin{abstract}
Via hydrodynamics preserving molecular dynamics simulations we study growth phenomena in a phase separating symmetric binary mixture model. We quench high-temperature homogeneous configurations to state points inside the miscibility gap, for various mixture compositions. For compositions around the symmetric or critical value we capture the  rapid linear viscous hydrodynamic growth due to advective transport of material through tube-like interconnected domains. The focus of the work, however, is on compositions close to any of the branches of the coexistence curve. In this case, the growth in the system, following nucleation of droplets of the minority species, occurs via coalescence mechanism. Using state-of-the-art techniques we have identified that these droplets, between collisions, exhibit diffusive motion. The value of the exponent for the power-law growth, related to this diffusive coalescence mechanism, as the composition keeps departing from the critical value, has been estimated. This nicely agrees with a theoretical number. These results are compared with the growth that occurs via particle diffusion mechanism, in non-hydrodynamic environment. 
\end{abstract}

\maketitle
\section{Introduction}
When quenched inside the coexistence region, a homogeneously mixed binary (A+B) system separates into phases that are rich in A and B particles \cite{binder_rep,binder_cahn,kash,ral,onuki,bray_adv,puri_book,majumder_2011,majumder_2013, mitchell,ahmed_2012}. This transformation occurs via formation and growth of domains of like particles. Such an evolution process is complex, during which the evolving structure exhibits interesting self-similar property \cite{binder_cahn,ral,onuki,bray_adv,puri_book}. The latter implies that the structures at two different times are same, except for the difference in size. As a consequence, one observes the scaling behavior \cite{bray_adv,binder_cahn}
\begin{equation}\label{scale_cf}
 C(r,t)\equiv\tilde{C}(r/\ell(t)),
\end{equation}
of the two-point equal time correlation function \cite{bray_adv,binder_cahn}
\begin{equation}\label{cf}
    C(r,t)=\langle\psi({\vec{r}}_1,t)\psi(\vec{r}_2,t)\rangle-\langle{\psi(\vec{r}_1,t)}\rangle \langle\psi(\vec{r}_2,t)\rangle,
\end{equation}
$\tilde{C}(x)$ being a master function that is independent of time.
In Eqs. (\ref{scale_cf}) and (\ref{cf}), $r=|\vec{r}_1 - \vec{r}_2|$, $\ell(t)$ represents the average size of domains at time $t$, and $\psi$ is a space and time dependent order parameter, defined as the local concentration difference between the two species \cite{ahmed_2012,dasgupta_2014,cates_2018}. Typically $\ell$ grows in a power-law fashion as \cite{bray_adv,binder_cahn,puri_book,ral,onuki}
\begin{equation}\label{powerlaw_l}
 \ell \sim t^\alpha.
\end{equation}

The growth exponent $\alpha$ depends upon several parameters \cite{binder_cahn,puri_book}. In non-hydrodynamic environment, one expects $\alpha=1/3$. This is referred to as the Lifshitz-Slyozov (LS) growth law \cite{majumder_2011,majumder_2013,ls,wagner,voorhees,huse,amar,marko,heerman} and is a result of diffusive transport of particles, via chemical potential gradient. The LS picture applies to phase separating solid mixtures and remains valid for critical as well as off-critical compositions \cite{majumder_2011,majumder_2013}, for the entire growth period. In fluids, however, hydrodynamics is important. There the mechanisms and exponents are different for the above two situations that give rise, respectively, to bicontinuous and disconnected droplet morphologies \cite{bs, binder_prb, siggia, furu_85, furu_87,miguel, farrel, oono_93, koga_93, tanaka_prl,tanaka_95,tanaka_96, laradji_prl, tanaka_97, lebo_prl_1997, kendon, charu,suman_epl, sr_2012, sr_soft_matt_2013, tanaka_2015,thakre,kabrede,koch,kumaran_98}. This is true for vapor-liquid as well as liquid-liquid transitions, in the former case density playing the role of composition. Below we briefly describe these in the liquid-liquid context.

In the case of a composition close to the critical value, say, a $50:50$ proportion of A and B particles for a symmetric model of mixture, growth occurs via transport of material through tube-like structure \cite{bray_adv,puri_book,ahmed_2012,furu_85,furu_87,siggia}. Overall growth in this situation is not described by a single exponent. At very early time LS picture remains valid \cite{bray_adv,puri_book}. Following this hydrodynamics becomes important, leading to a crossover of the exponent to $\alpha=1$, in space dimension $d=3$, which is referred to as the viscous hydrodynamic growth \cite{bray_adv,puri_book,ahmed_2012,siggia,kendon,sr_2012,sr_soft_matt_2013}. At an even later time a further crossover occurs to a smaller value, viz., $\alpha=2/3$, known as the inertial hydrodynamic exponent \cite{bray_adv,puri_book}. For a composition close to any of the branches of the coexistence curve, on the other hand, the late time growth, in an hydrodynamic environment, may occur via coalescence of disconnected droplets that consist primarily of particles of the minority phase \cite{siggia,bs,binder_prb,sr_2012,sr_soft_matt_2013,tanaka_prl,tanaka_96,tanaka_97,tanaka_95,kumaran_98}. This we discuss below. 

For diffusive motion of the droplets, between collisions, expected for liquid mixtures, because of high density background phase, a theory for growth was proposed by Binder and Stauffer (BS) \cite{bs,binder_prb,siggia}. In this case, solution of the dynamical equation \cite{bs}
\begin{equation}\label{bs}
    \frac{\textrm{d}n}{\textrm{d}t}=-D\ell n^2,
\end{equation}
for droplet density $n\, (\propto {1}/{{\ell}^d}$), provides $\alpha={1}/{d}$. In Eq. (\ref{bs}), $D$ is a diffusion constant, having dependence upon $\ell$. It is expected that $D\ell$ will remain a constant, during the growth period, in accordance with the generalized Stokes-Einstein-Sutherland \cite{hansen,das_jcp_2007,brady} relation. In $d=3$, the BS value is same as the LS exponent. The difference in the mechanisms is expected to be captured in the amplitudes of growth, this being larger for the BS case. The ratio of the amplitudes for the two mechanisms is supposed to follow the relation \cite{tanaka_97,sr_2012}
\begin{equation}\label{ratio}
    \frac{A_{BS}}{A_{LS}}=K\phi^{1/3}, 
\end{equation}
where $K$ is a constant ($\simeq 6$) and $\phi$ is the volume fraction of the minority species in the mixture. 

In this work, while presenting results for a wide range of compositions, we primarily address the issue of growth via diffusive coalescence in a phase separating binary fluid mixture. We have performed molecular dynamics \cite{frenkel,allen_tild} (MD) simulations. In our canonical ensemble simulations the temperature is controlled via the Nos\'e-Hoover thermostat (NHT) \cite{frenkel,nose,hoover,hoover_book}. The latter is known to preserve hydrodynamics. The obtained results were analyzed via various advanced methods to arrive at conclusions on the growth and mechanism. 

\section{ Model and Methods}
In our model system, two particles, located at $\vec{r}_i$ and $\vec{r}_j$, with $r=|\vec{r}_i-\vec{r}_j|$, interact via the potential \cite{allen_tild}:
\begin{equation}\label{LJ}
U(r)= V(r)-V(r_{c})-(r-r_{c})\left[\frac{dV(r)}{dr}\right]_{r=r_{c}},
\end{equation}
for $r<r_c$, the latter being a cut-off distance. In Eq. (\ref{LJ}), $V(r)$ is the standard Lennard-Jones (LJ) potential \cite{allen_tild}
\begin{equation}\label{LJ_ori}
 V(r)=4\epsilon_{\alpha\beta}\left[\left(\frac{\sigma_{\alpha\beta}}r\right)^{12}-
 \left(\frac{\sigma_{\alpha\beta}}r\right)^{6}\right].
\end{equation}
The cut and shift, implemented via the second term on the right hand side of Eq. (\ref{LJ}), for the sake of computational convenience, leaves the force at $r=r_c$ discontinuous. This was corrected via the introduction of the last term \cite{allen_tild} in the same equation. In Eq. (\ref{LJ_ori}), $\epsilon_{\alpha\beta}$ and $\sigma_{\alpha\beta}$, $\alpha,\beta \in [{\textrm{A,B}}] $, represent the interaction strengths and diameters, respectively, for various combinations of particles. In this work we have chosen \cite{das_prl,das_jcp}
\begin{equation}\label{sigma}
\sigma_{AA}=\sigma_{BB}=\sigma_{AB}=\sigma,   
\end{equation}
and
\begin{equation}\label{epsi}
\epsilon_{AA}=\epsilon_{BB}=2\epsilon_{AB}=\epsilon. 
\end{equation}
We have also set mass ($m$) of all the particles to be the same.

For this {\it{symmetric}} model, the phase diagram, in $d=3$, is accurately known for (number) density of particles $\rho=1$. It is, of course, expected \cite{das_prl,das_jcp,roy_epl} that the critical concentration, for any of the species, will be $x^c_{\alpha}=1/2$. The critical temperature, $T_c$, was estimated to be $\simeq 1.421 \epsilon/k_B$, where $k_B$ is the Boltzmann constant \cite{das_prl,das_jcp,roy_epl}. In the following we set $\sigma$, $\epsilon$, $m$ and $k_B$ to unity. In this work we study the kinetics by quenching homogeneous configurations, prepared at a high temperature, to the temperature $T=1$. The compositions are chosen in such a way that the final state points fall inside the miscibility gap \cite{das_prl,das_jcp,roy_epl}. 

We have performed MD simulations \cite{frenkel} in periodic cubic boxes of linear dimension $L$, the latter being measured in units of $\sigma$. As already stated, the temperature was controlled via the application of a NHT \cite{frenkel,nose,hoover,hoover_book} that is known to preserve hydrodynamics well. For comparative purpose, towards the end, we have also presented results that were obtained via the application of a stochastic thermostat, viz., the Andersen thermostat (AT) \cite{frenkel,and}. 

In our MD method we have used the Verlet velocity integration scheme \cite{frenkel,allen_tild}, with time discretization step $\Delta t=0.001 \tau$. Here $\tau$ $(=\sqrt{m\sigma^2/\epsilon})$ is our LJ unit of time, which is unity because of the above mentioned choices of $\epsilon$, $\sigma$ and $m$. All our quantitative results are obtained after averaging over runs with a minimum of $50$ independent initial configurations. Unless otherwise mentioned, all the simulations were performed with $L=64$ and NHT. 

We have calculated lengths from the decay of the correlation function as
\begin{equation}\label{dl_from_cf}
    C(r=\ell,t)=c,
\end{equation}
by fixing $c$ to the first zero of the correlation function. It is worth mentioning here, for conserved order-parameter dynamics, the class to which the present problem belongs, $C$ exhibits damped oscillation around zero. This we will see in the next section. For the calculation of $C(r,t)$ we have mapped the continuum configurations to the ones on a simple cubic lattice \cite{majumder_2011}. The length was also obtained by direct identification of the droplets \cite{hoshen} and counting numbers of particles within those \cite{roy_jcp}. The latter, of course provides volume, from which average length can be trivially obtained following calculation of the average volume via the first moment of a distribution. Results from different methods are essentially proportional to each other, differing by constant factors.

For disconnected morphology, the droplet identification \cite{hoshen} is important for the purpose of confirmation of the mechanism as well, e.g., via the calculation of mean-squared-displacement (MSD) of the centres of mass (CM) of the droplets. Note that for $N_p$ particles belonging to a particular droplet the centre of mass is calculated as \cite{hansen}
\begin{equation}\label{r_cm}
    \vec{R}_{CM}(t)=\frac{1}{N_p}\sum_{i=1}^{N_p} \vec{r}_i(t).
\end{equation}
The MSD is then obtained from the formula \cite{hansen}
\begin{equation}\label{msd}
    {\textrm{MSD}}=\langle (\vec{R}_{CM}(t')-\vec{R}_{CM}(0))^2\rangle.
\end{equation}
Here $t'$ is a time that is shifted with respect to the time at the beginning of an observation.

\section{ Results} 
\begin{figure} 
\centering
\includegraphics*[width=0.4\textwidth]{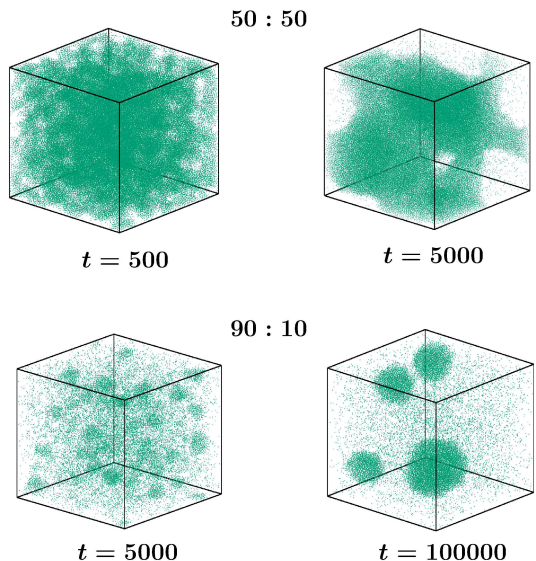}
\caption{\label{fig1}Snapshots, that were recorded during the molecular dynamics simulations, following quenches of high temperature homogeneous configurations to $T=1$, are shown for $50:50$ (upper panel) and $90:10$ (lower panel) compositions of A and B particles. Only the locations of the B particles are marked. For each of the compositions, frames from two different times are included.}
\end{figure}
In Fig. \ref{fig1} we show snapshots taken during the evolutions of two typical homogeneously mixed configurations towards respective equilibriums, following quenches inside the coexistence curve. For $50:50$ composition, it is appreciable that the morphology consists of interconnected tube-like domains. For the asymmetric composition, disconnected droplet morphology is clearly identifiable. In the latter case the growth seems to be much slower. Our objective here is to provide a composition dependent quantitative picture and obtain an understanding on the pathway to the equilibrium for the disconnected case. 

\begin{figure}
    \centering
    \includegraphics*[width=0.4\textwidth]{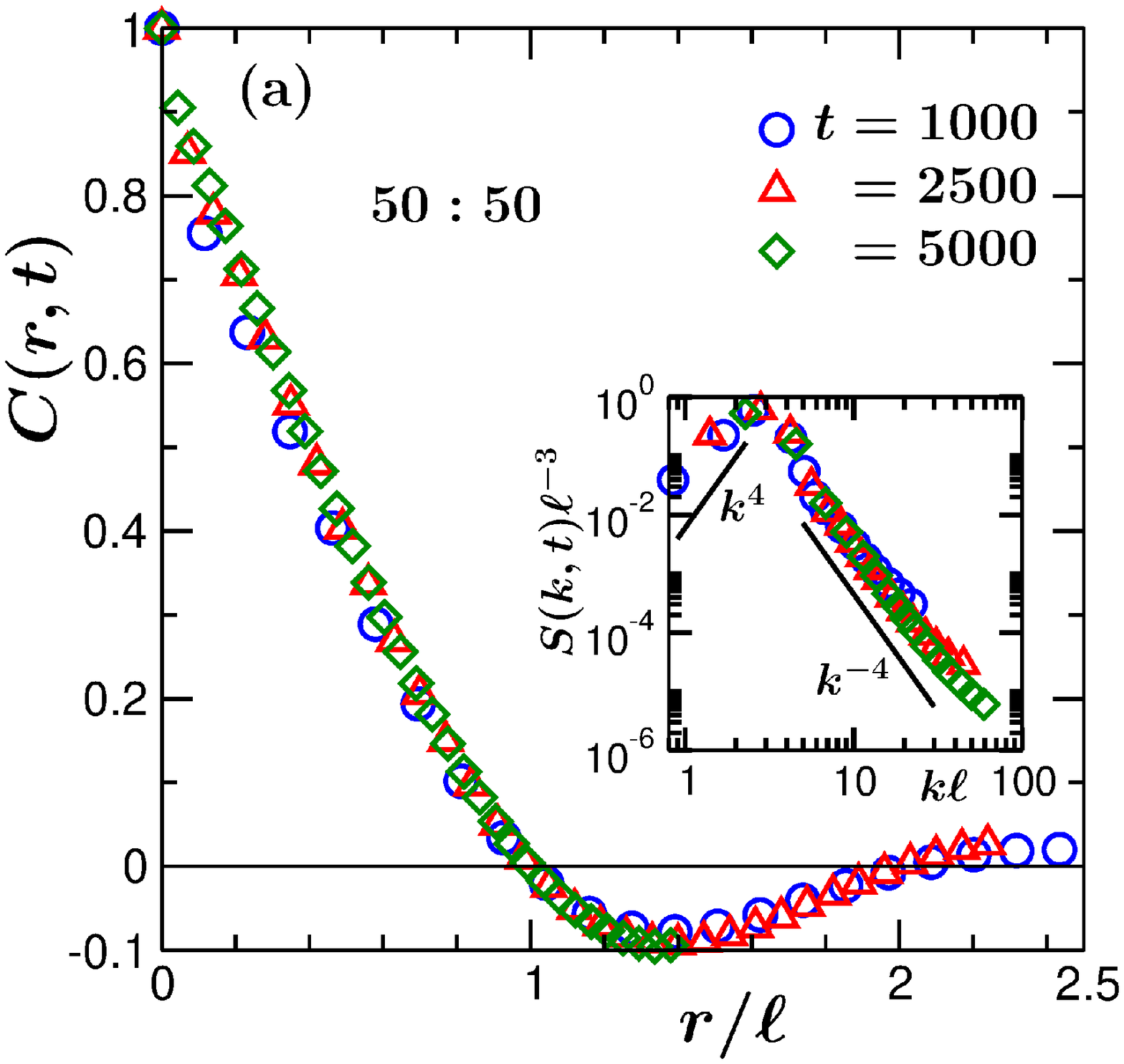}
    \vskip 0.3cm
    \includegraphics*[width=0.4\textwidth]{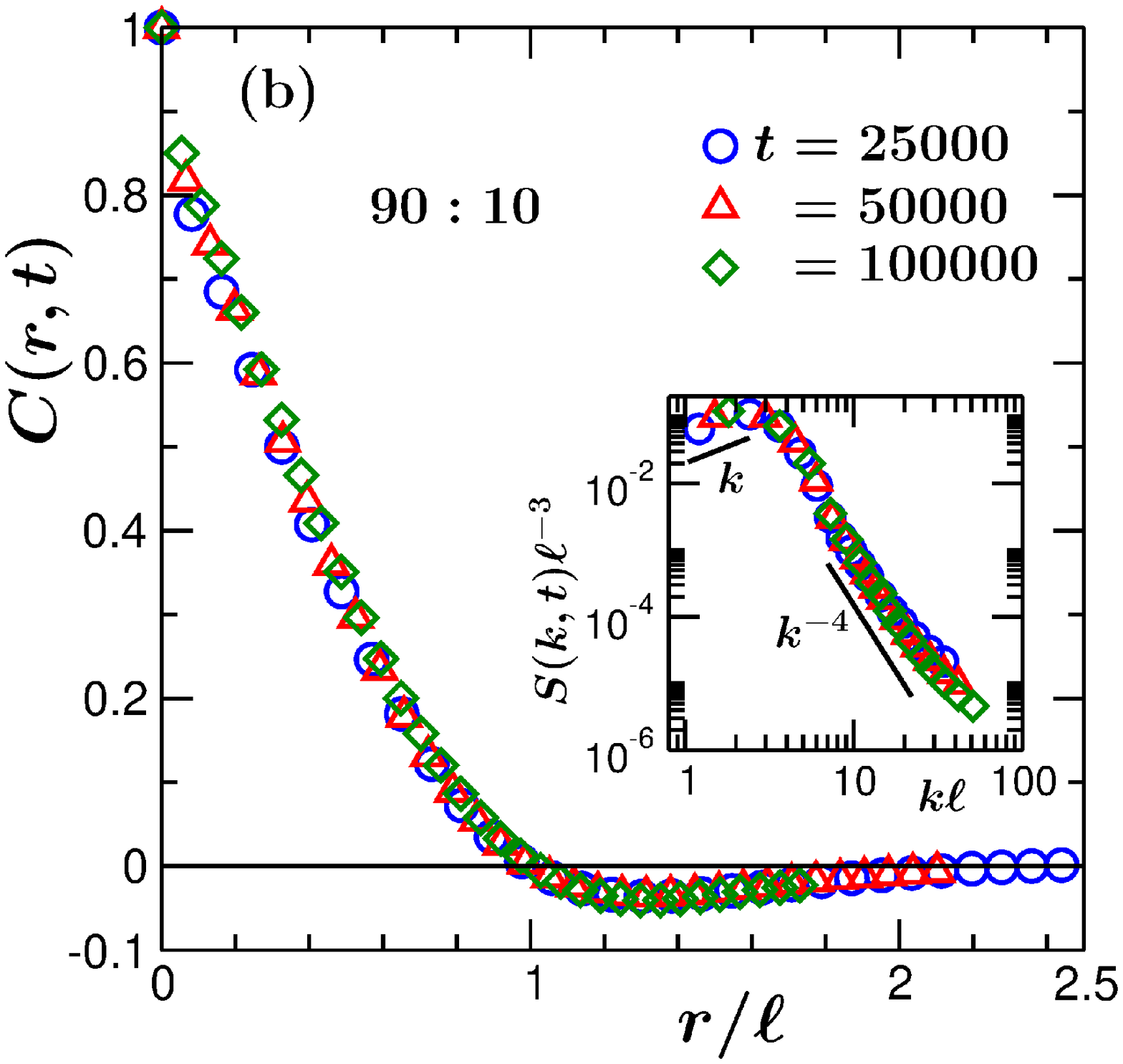}
    \caption{\label{fig2} (a) Two-point equal time correlation functions, $C(r,t)$, are shown, from a few different times, versus the scaled distance $r/\ell(t)$, for $50:50$ composition. In the inset we show the analogous scaling plots for the structure factor,  $S(k,t)$, $k$ being the wave number. The solid lines represent power laws. (b) Same as (a) but here the composition is $90:10$.}
\end{figure}
In Fig. \ref{fig2} we show scaling exercise \cite{bray_adv} for the correlation function. There $C(r,t)$ is plotted versus $r/\ell$. Results from a few different times, for two compositions, have been included. Part (a) contains results for the symmetric composition and the results for $90:10$ composition are included in part (b). Nice collapse of the data sets imply self-similarity of growth in both the cases. For conserved order-parameter dynamics one expects damped oscillation of $C(r,t)$ around zero. This is clearly visible here. In the asymmetric composition case the minimum is expectedly quite shallower compared to the symmetric or critical ($50:50$) composition case \cite{das_puri_pre,majumder_2013,paul_pre}.

\begin{figure}
\centering
\includegraphics*[width=0.4\textwidth]{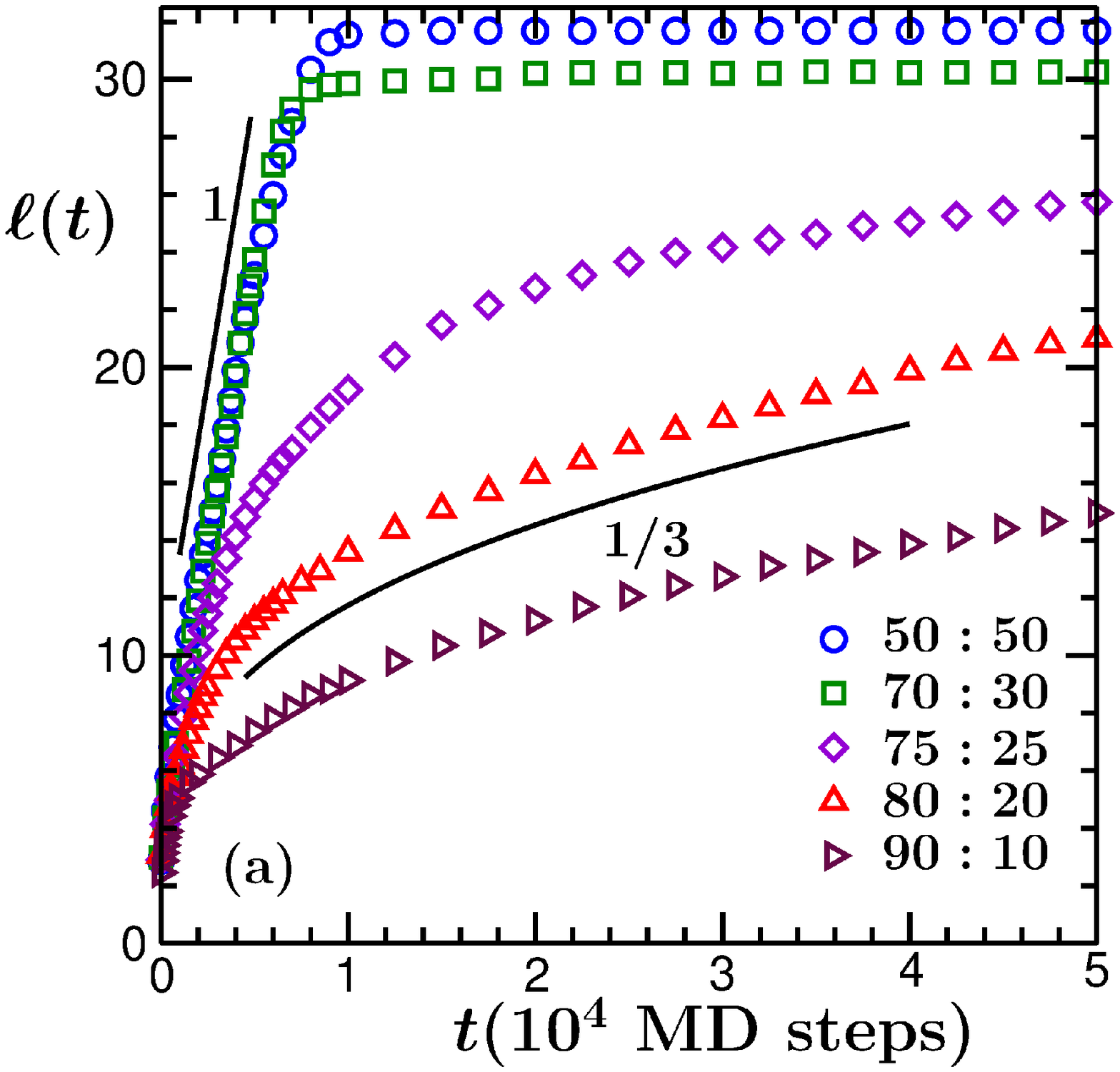}
\vskip 0.3cm
\includegraphics*[width=0.4\textwidth]{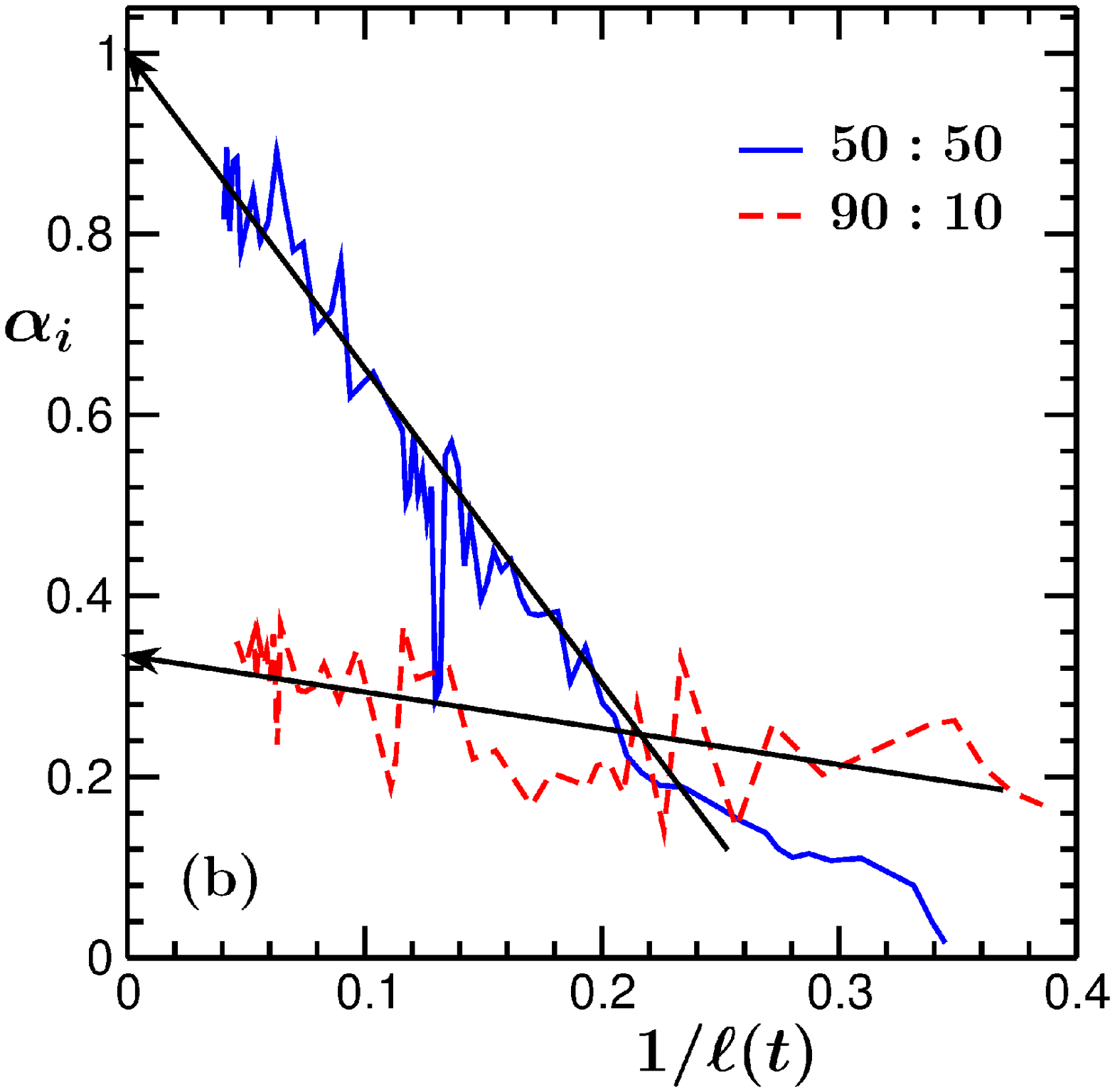}
\caption{\label{fig3}(a) Average domain lengths, $\ell(t)$, are shown with the variation of time, for quenched systems having different compositions of A and B particles. The solid lines represent power-laws with mentioned exponents. (b) Instantaneous exponents are shown as a function of $1/\ell$, for the compositions $50:50$ and $90:10$. The arrow-headed lines there are guides to the eyes.}
\end{figure}
\begin{figure}
\centering
\includegraphics*[width=0.4\textwidth]{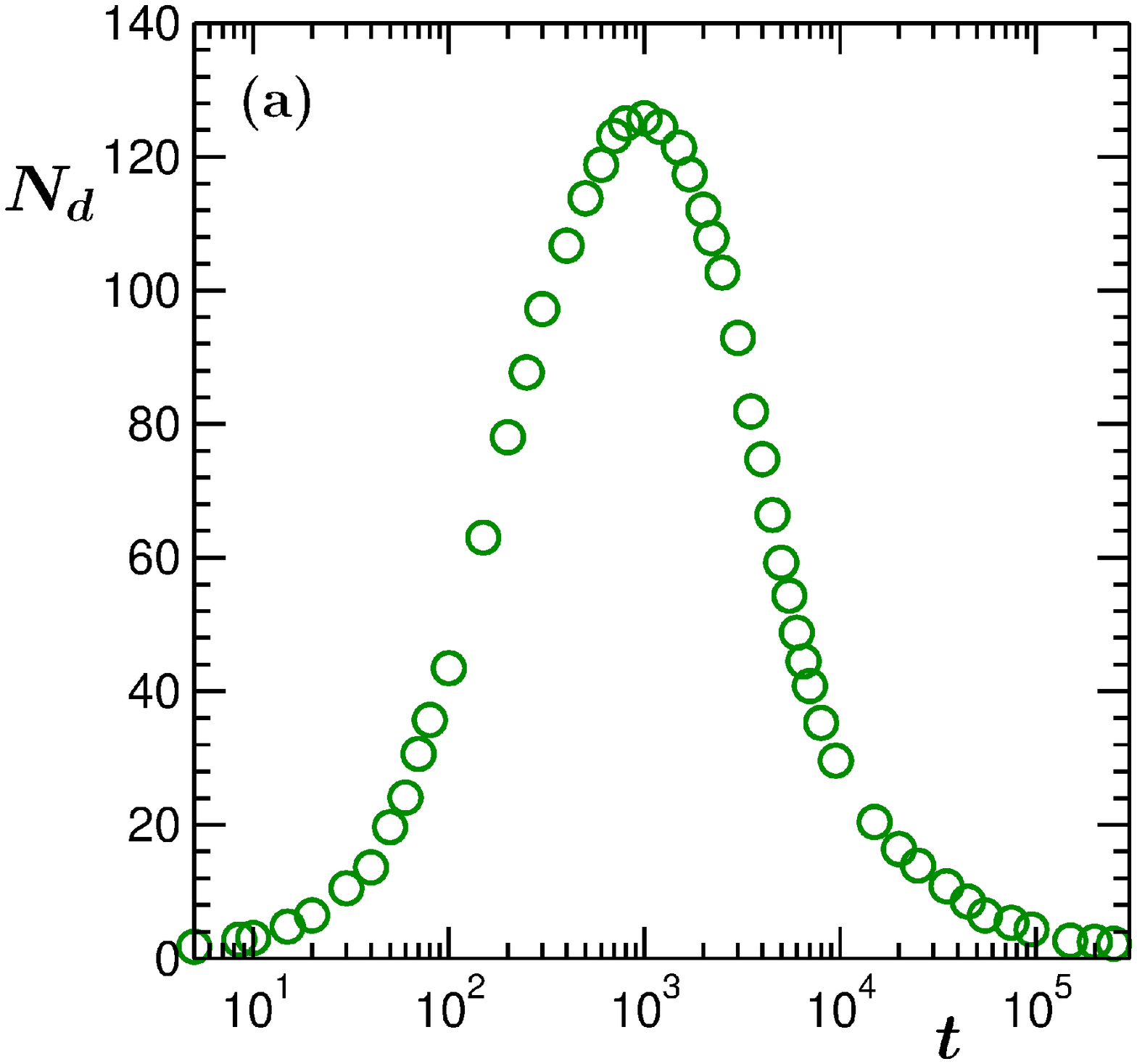}
\vskip 0.3cm
\includegraphics*[width=0.44\textwidth]{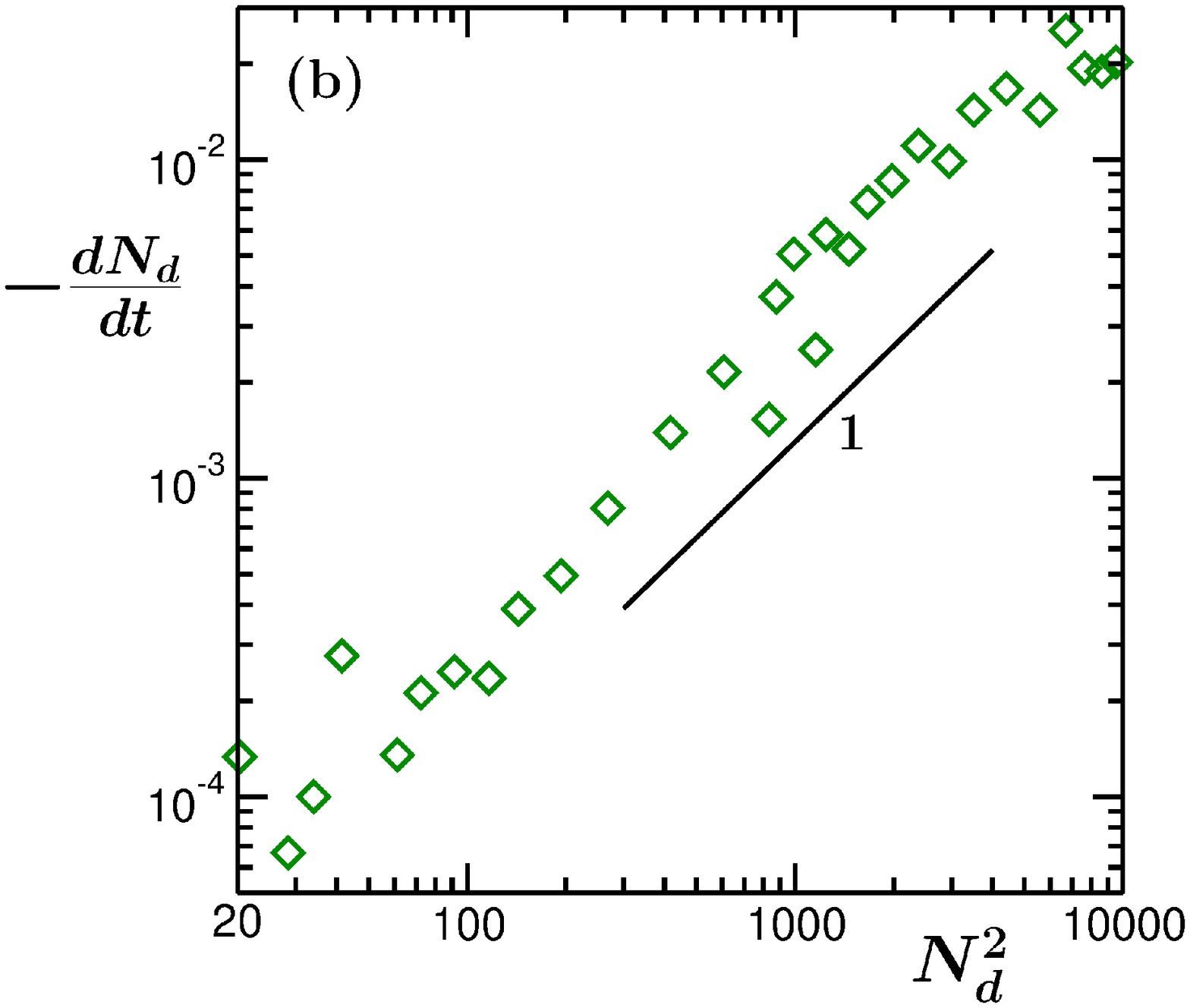}
\caption{\label{fig4}(a) Here we have plotted the number of droplets, consisting primarily of the particles of the minority phase, as a function of time, on a semi-log scale, for $90:10$ composition. (b) A plot of $-{\textrm{d}N_d/\textrm{d}t}$ versus ${N_d}^2$, on a log-log scale, corresponding to the plot in (a). The solid line is a power-law with exponent $1$.}
\end{figure}
\begin{figure}
\centering
\includegraphics*[width=0.4\textwidth]{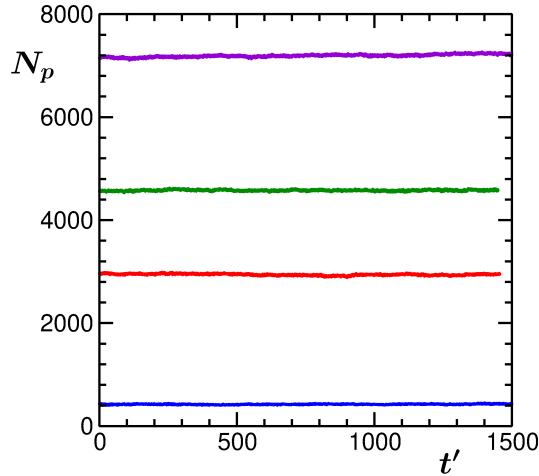}
\caption{\label{fig5} Numbers of particles, $N_p$, in a few droplets, are shown as a function of the translated time $t'=t-t_0$, $t_0$ being the beginning of an observation. These results are for the composition $90:10$ and $L=48$. During the presented periods the considered droplets did not undergo collision with other droplets.}
\end{figure}
\begin{figure}
    \centering
    \includegraphics*[width=0.4\textwidth]{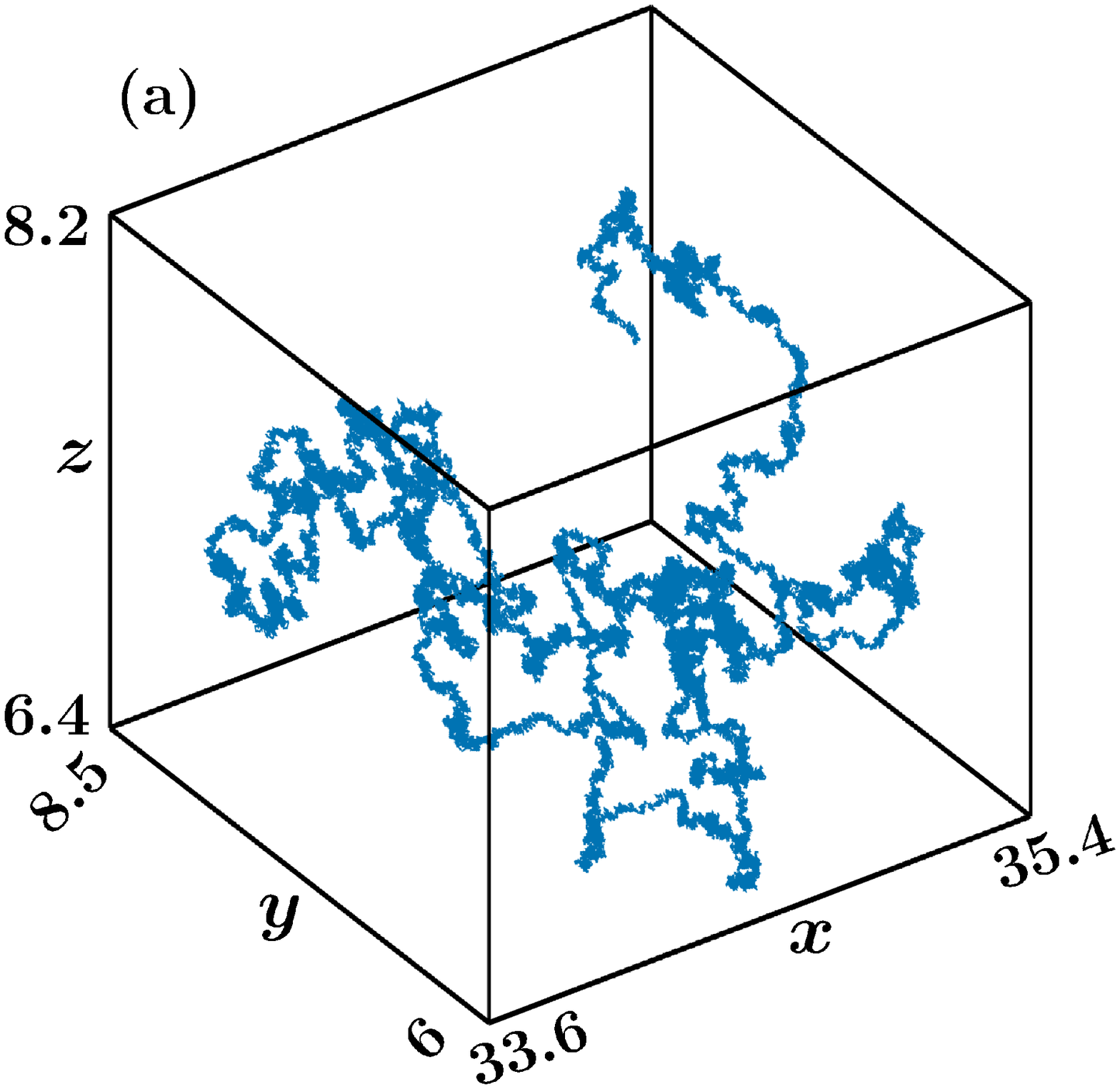}
    \vskip 0.3cm
    \includegraphics*[width=0.4\textwidth]{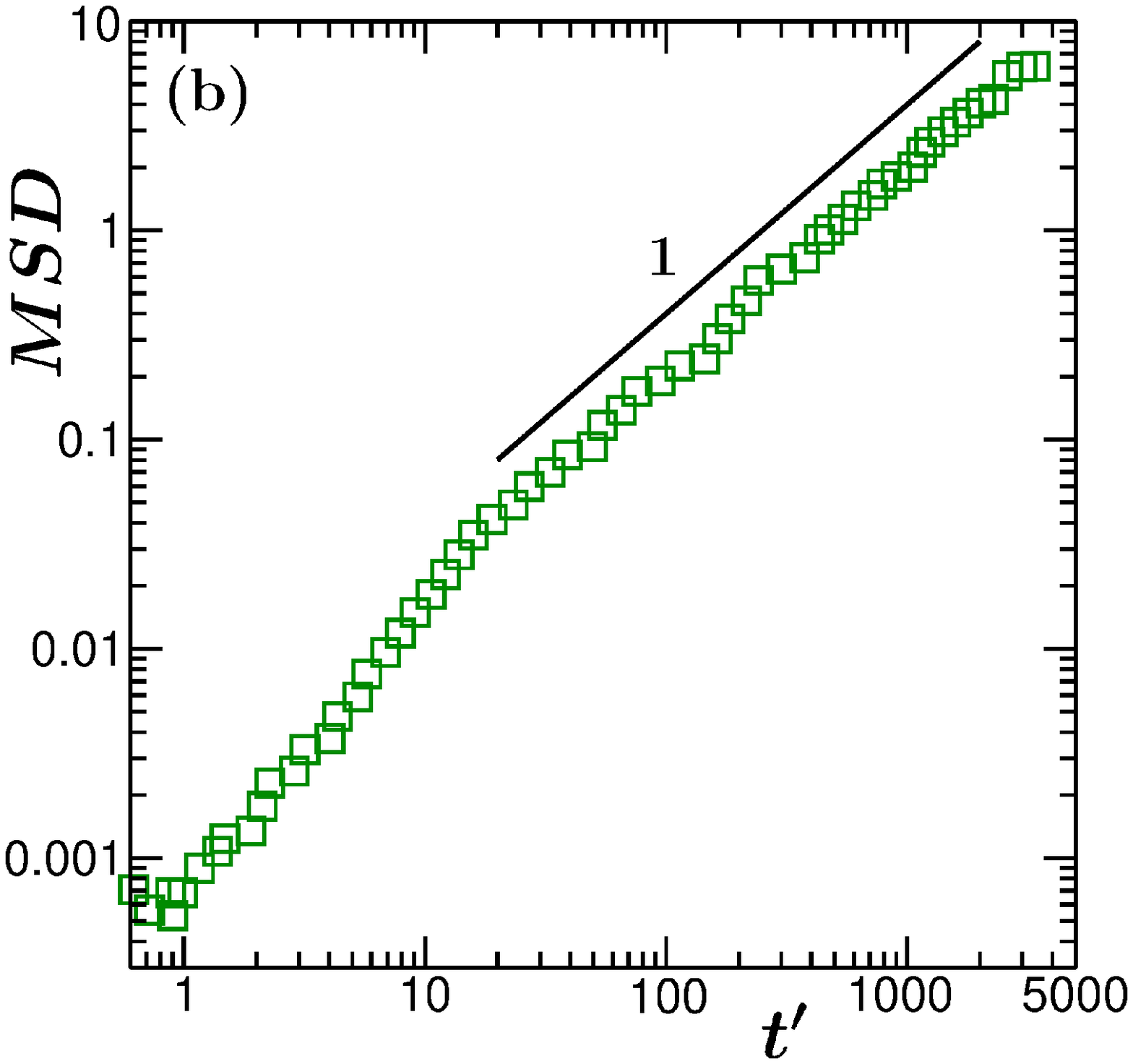}
    \caption{\label{fig6} (a) Here we show the trajectory of the centre of mass of a typical droplet. (b) Log–log plot of the mean-squared-displacement (MSD) of a droplet, as a function of the shifted time $t'$. During this period the droplet did not encounter any collision with other droplets. The solid line represents the diffusive displacement. These results are for the composition $90:10$.}
\end{figure}
\begin{figure}
\centering
\includegraphics*[width=0.4\textwidth]{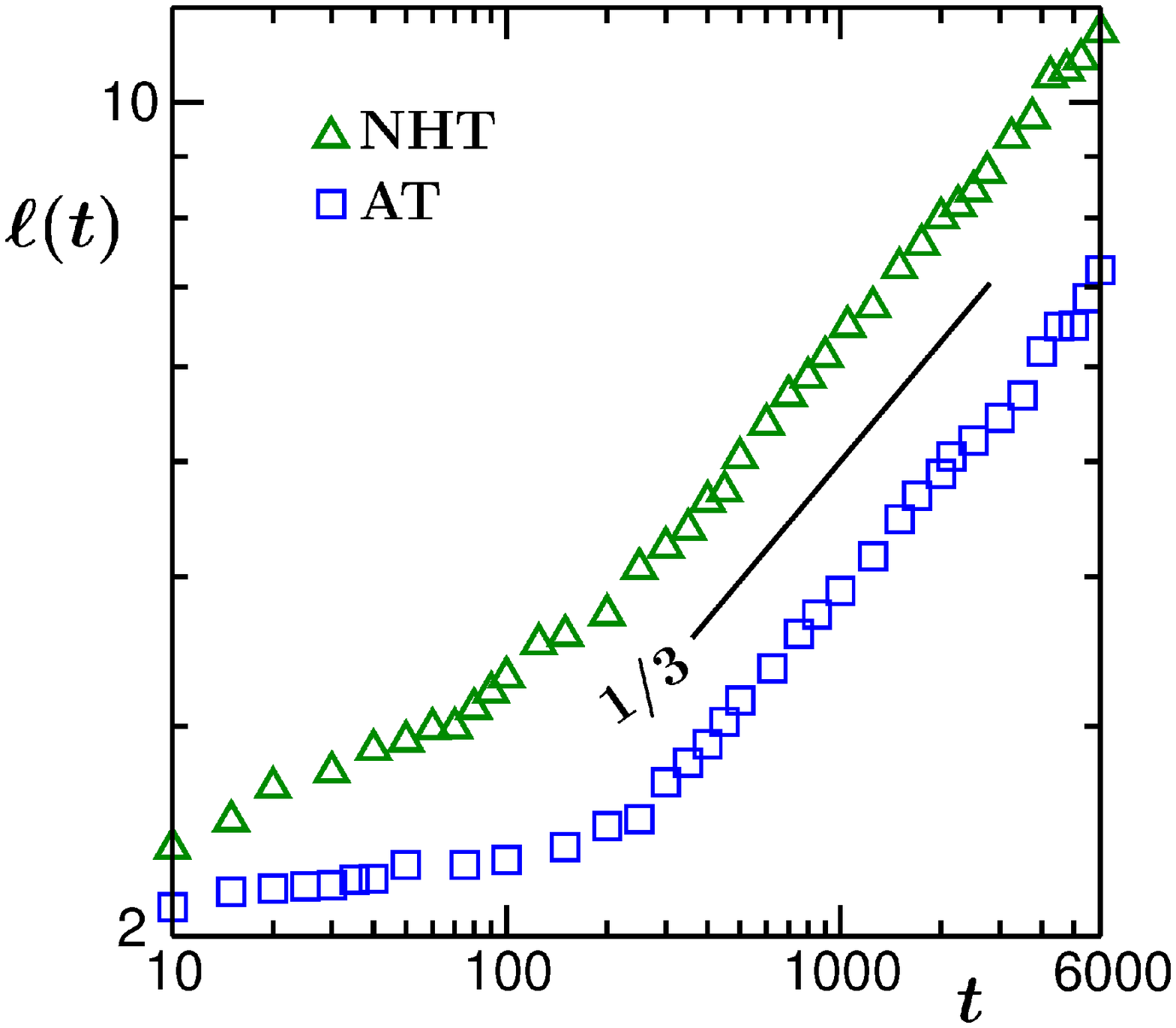}
\caption{\label{fig7} Plots of $\ell$ versus time, on a log-log scale. Here we have compared the results obtained via the applications of NHT with those gathered by using AT. These results are for the composition $80:20$. The solid line is a power-law with exponent $1/3$.}
\end{figure}
In the insets of Fig. \ref{fig2} we have shown the scaling plots of $S(k,t)$, the structure factor, a quantity that is of direct experimental relevance \cite{bray_adv}. This is the Fourier transform of $C(r,t)$. The expected scaling form for this quantity is \cite{bray_adv} 
\begin{equation}\label{scaling_sf}
    S(k,t) \equiv {\ell^d} \tilde{S}(k\ell),
\end{equation}
where $\tilde{S}(y)$ is another time independent master function.
Clearly, good collapse of data from different times is visible. The small $k$ behavior is consistent with \cite{ahmed_2012,paul_pre} $k^4$, for the symmetric case, referred to as the Yeung's law \cite{yeung}. The result for the asymmetric case, however, is at deviation \cite{paul_pre} with the Yeung's law. The large $k$ behavior is consistent with a power-law having an exponent $-4$, in both the cases. This is the expected Porod law \cite{porod,oono,bray_puri} in $d=3$, for a scalar order-parameter, and is an outcome of scattering from sharp interfaces. The deviations that are observed can be due to the interfacial roughness that is appreciable from the noise that is noticed in the snapshots of Fig. \ref{fig1}. Such noise can be gotten rid of via choices of lower quench temperatures. This will, however, lead to slower dynamics, preventing us from accessing the desired growth regimes over appreciable periods of time. Even crystallization is a possibility. In our study, for the purpose of analysis, this noise was largely eliminated via the application of a majority rule \cite{majumder_2011}. Before moving to the discussion on growth, we mention that $\ell$ can be estimated from the first moment of $S(k,t)$ as well \cite{bray_adv}.

In Fig. \ref{fig3} (a) we have shown $\ell$ versus $t$ plots for several different compositions. For compositions at or close to the symmetric value, intermediate time behavior, over long periods, is linear, consistent with the expectation for viscous hydrodynamic growth. The saturations at late times are due to finite size of the systems. Late time behavior, for compositions far away from the critical value, is consistent with $\alpha=1/3$. This is expected for the BS \cite{bs,binder_prb,siggia} mechanism, in $d=3$. For a more convincing confirmation, of the values of the exponent, we have calculated the instantaneous exponent \cite{majumder_2011,huse,amar}, $\alpha_i= \textrm{dln}{\ell}/\textrm{dln}{t}$. This quantity is shown in Fig. \ref{fig3}(b), as a function of $1/\ell$. The convergences, in the $\ell \rightarrow \infty$ limit, to $\alpha=1$ and $\alpha=1/3$, can be appreciated. 
For $\ell<\infty$, smaller values of $\alpha_i$, compared to the expected ones, can be attributed to the presence of non-zero offsets at the beginning of a scaling regime. If such initial length has a value $\ell_0$, $\alpha_i$ is expected to exhibit the behavior \cite{majumder_2011,amar}
\begin{equation}\label{ie_corr}
    \alpha_i=\alpha\left[1-\frac{\ell_0}{\ell}\right].
\end{equation}
Clearly, for both the compositions, data in Fig \ref{fig3} (b) are consistent with Eq. (\ref{ie_corr}). In the rest of the paper we focus only on the discontinuous morphology, for the representative case of $90:10$ composition. Having identified the exponent for the power-law growth, in the following we present results that will ascertain that the growth indeed occurs via diffusive coalescence mechanism, when hydrodynamics is preserved. 

In Fig. \ref{fig4} (a) we present a plot for the number of droplets as a function of time. The early part is dominated by nucleation. The late time decay is due to growth. In part (b) of this figure we have shown a plot of $-{dN_d/dt}$ versus ${N_d}^2$. There the focus is on the late time growth part. Thus, we have shown data from $t=5000$ onward. The linear behavior on a double-log scale indicates a power-law. We expect \cite{sr_soft_matt_2013} an exponent $1$ for diffusive coalescence --- see Eq. (\ref{bs}) and related discussion. That indeed is observed. This also indirectly validates the Stokes-Einstein-Sutherland \cite{hansen,das_jcp_2007,brady} relation in this extended context.

In Fig. \ref{fig5} we show numbers of particles in several droplets, with the variation of time $t'$ that is calculated from the beginning of an observation. During the presented periods these droplets did not collide with any other droplets. In each of the cases the value of $N_p$ remains practically constant. This again suggests that LS-like particle diffusion mechanism \cite{ls} is playing negligible role in the growth. The number of droplets is decreasing, as seen in Fig. \ref{fig4}, due to coalescence. 

In Fig. \ref{fig6} we have shown results related to the motion of the centres of mass of the droplets. In part (a) a trajectory of the CM of a typical droplet is seen. Random motion of the droplet is visible. In part (b) we have shown MSD versus time plot for such a droplet. Clearly diffusive displacement is visible, at late times.

The above results suggest that the growth is occurring via the diffusive coalescence mechanism. In Fig. \ref{fig7} we provide further information. There we have compared growth plots obtained for NHT and AT. Hydrodynamic effect is not expected for AT. Thus, NHT should provide faster evolution, even though the exponent in both the cases should be the same. Plots of Fig. \ref{fig7} are consistent with this picture. Certainly $A_{BS}/A_{LS}$ is $> 1$. However, a proper match with the theoretically expected number is not obtained \cite{sr_2012}. This is due to the fact that even though NHT provides hydrodynamics, a perfect match of transport with natural system can not be expected, unless various thermostat parameters are appropriately tuned.

\section{Conclusion}
We have studied kinetics of phase separation in a high density symmetric binary (A+B) fluid model \cite{allen_tild,das_prl,das_jcp,roy_epl}. We have considered mixture compositions symmetric as well as close to one of the branches of the coexistence curve. While around the symmetric composition a bicontinuous nonequilibrium domain morphology is obtained, for compositions close to a coexistence curve the domain morphology consists of disconnected droplets of the minority phase in the background of a sea consisting of particles of the majority species. The symmetric case is much studied and in agreement with the previous studies \cite{ahmed_2012,furu_85,furu_87,siggia} we observe linear viscous hydrodynamic evolution within the permissible simulation length and time scales.

We have performed molecular dynamics simulations for our study. To capture hydrodynamics, in our canonical ensemble simulations, we have used a thermostat that is known for its capability of conservation \cite{frenkel} of local momentum, etc. Main focus is on the disconnected morphology. These results were compared with those obtained via the application of a stochastic thermostat \cite{frenkel}.

We observe that the droplets are not static in hydrodynamic environment. Via the calculation of mean-squared-displacements of the centres of mass, we show that these exhibit diffusive motion. Due to sticky collisions among these droplets, the number density of these objects in the system decreases, thereby the characteristic length scale increases. For this diffusive coalescence mechanism we have accurately estimated the exponent for the power-law growth. This in good agreement with a theoretical expectation \cite{siggia,bs,binder_prb}.

The picture described above is different from that is provided by the results obtained in a stochastic situation. In this case the droplets are practically static and growth occurs via particle diffusion, as in solid mixtures \cite{ls,majumder_2011,majumder_2013,huse,amar} . 
However, the exponent remains same in both the cases. We have estimated the ratio of the growth amplitudes in the two cases. This suggests that the hydrodynamic growth is faster.

In future it will be interesting to study the aging properties \cite{fisher_huse} of such off-critical binary mixtures. It will be important to compare the results obtained with and without hydrodynamics. 

\section{Acknowledgement}
The authors acknowledge computer times at the National Supercomputing Mission
located at JNCASR.

\end{document}